# Near-field Nanoscopy of Thermal Evanescent Waves on Metals

S. Komiyama,[1,*] Y. Kajihara,[1,2] K. Kosaka,[1] T. Ueda,[1] and Zhenghua An[3]

[1] Department of Basic Science, The University of Tokyo, Komaba 3-8-1, Meguro-ku, Tokyo, 153-8902, Japan.
[2] Institute of Industrial Science, The University of Tokyo, Komaba 4-6-1, Meguro-ku, Tokyo, 153-8505, Japan.
[3] State Key Laboratory of Surface Physics, Department of Physics and Collaborative Innovation Center of Advanced Microstructures, Fudan University, Shanghai 200433, PR China
 *e-mail: csusukom@mail.ecc.u-tokyo.ac.jp

Thermally excited electromagnetic evanescent waves on material surfaces are of great importance, not only for understanding fundamental surface-related properties such as friction forces, Casimir forces, and near-field heat transfer, but also for exploiting novel technologies including magnetic recording, near-field thermophotovoltaics and lithography. On metal surfaces, relevance of surface plasmon polaritons (SPlPs), coupled with collective motion of conduction electrons, is of strong interest, but has not been clarified so far. Here, using a passive near-field microscope with unprecedented high sensitivity, we unveil detailed nature of thermal evanescent waves. Our experiments unambiguously indicate that the thermal waves are dominated by fluctuating electromagnetic fields with extremely short in-plane wavelengths and extremely short out-of plane decay lengths, which makes unlikely significant contribution of SPlPs. This work, contributing to deeper understanding of surface-related phenomena of metals, provides the basis for rational design of a variety of devices exploiting the heat-radiation coupling on nanometer scales.

Existence of intense thermal electromagnetic evanescent waves[1] on metals and polar dielectrics has been experimentally proved by a series of measurements, in which radiative heat conduction between two bodies dramatically increases in the near field.[2-8] In the case of polar dielectrics, it is established that the evanescent waves are dominated by surface phonon polaritons (SPhPs), which give a quasi monochromatic sharp peak in the spectrum.[5-8] By analogy to the SPhP in dielectrics, one would expect that surface plasmon polaritons (SPlPs)[9-11] play a key role in metals. Experimentally, fringe-like interference patterns were reported in the images of passive scattering–type scanning near-field optical microscope on heated Au layers and relevance of thermally excited SPlPs was suggested.[12] Our work is a thorough reexamination of thermal electromagnetic evanescent waves on metals with an improved passive s-SNOM system ($\lambda_0 = 14.5 \pm 0.7$ μm). Owing to the ultra-high sensitivity and unprecedented high spatial resolution (20 nm) of the system, our measurements unveil detailed new features of the waves that are largely different from those reported earlier:[12] Thermal evanescent fields are utterly dominated by fluctuating electromagnetic fields with extremely short in-plane wave lengths $\lambda_{//} \ll \lambda_0$ and extremely short out-of-plane decay lengths $l_z \ll \lambda_0$, and no interference patterns ascribable to SPlPs are seen in the near-field image. All of the experimental findings, including temperature dependence, are consistently explained by existing theories,[1,13,14] and establish deeper understanding of thermally excited electromagnetic waves on metals.

Experimental challenge in the study of thermal evanescent waves on metals stems from the fact that thermal energy density spectrum in the near field is monotonous without a characteristic peaked structure.[1] This is different from polar dielectrics, in which sharp radiation spectrum of SPhPs can be exploited as a distinct signature for identifying thermal evanescent waves.[15] For characterizing the waves on metals, therefore, frequency spectroscopy is inadequate and the study of wave vectors $\mathbf{k} = (\mathbf{k}_{//}, k_\perp)$ relevant to the waves at a given frequency is essential, where the wave vector, $\mathbf{k}_{//}$, parallel to the surface corresponds to the inverse in-plane wavelength, $\lambda_{//} = 2\pi/k_{//}$ with $k_{//} = |\mathbf{k}_{//}|$ and the wave number, $k_\perp$, in the direction normal to the surface, being imaginary, implies the inverse out-of-plane decay length, $l_z = 1/ik_\perp$ of the wave. Here $k_{//}$ for evanescent waves is larger than the wave number $k_0 = 2\pi/\lambda_0 = \omega/c$ in vacuum (ω; the angular frequency, c; the light velocity in vacuum) because $k_{//}^2 + k_\perp^2 = k_0^2$ is satisfied. To have experimental access to $\mathbf{k}$, we have developed an ultra-highly-sensitive passive scattering-type scanning near-field optical microscope (s-SNOM) in the long-wavelength infrared (LWIR) region (Fig. 1). With the improved s-SNOM system, wave vectors involved in the waves have been quantitatively clarified and deeper understanding of thermal evanescent waves on metals has been thereby achieved.

## Results
**Sensitive passive s-SNOM and extraction of near-field signals.** Special attention should be paid in the high-resolution passive s-SNOM measurements of metals, in which thermal evanescent waves, generated without external illumination, is scattered by a sharp probe tip to be detected. The first point to be noted is that the relevant energy flux of the tip-scattered near-field thermal radiation is so small as to be not readily detectable by commercially available detectors. In polar dielectrics the energy density of thermal evanescent waves is peaked in the narrow spectral region of SPhPs, but such enhancement is absent and the average radiation intensity is by orders of magnitude lower in metals. In several existing experiments,[12,15] samples are heated up for intensifying the signal, but heating up the sample can cause serious complication in measurements as described below. In this work, in order to obtain near-field signals without heating up the sample, we apply an ultra-highly sensitive LWIR detector, called CSIP,[16] and incorporate it in a specially designed home-made LWIR confocal microscope (see Methods).[17]

The second point to be noted is that the signal of small energy flux lies buried in much stronger far-field background radiation, which primarily arises from the spontaneous emission from the sample surface and from the reflection of ambient radiation at the sample surface, both in the focus spot of the confocal microscope. Whereas the near-field signal component is maximized in our measurement system, it is only a fraction about $10^{-3} \sim 10^{-4}$ of the far-field background. This is a serious issue for the study of metals because the spectrum of near-field evanescent radiation cannot be readily distinguished from that of the far-field background radiation. It is, therefore, essential to extract the near-field component in an reliable method, and to make sure that the near-field signal component is strictly extracted, by carefully examining the consistency of obtained results. For this sake, we oscillate the probe tip height independently of the control of the atomic force microscope (AFM) (see Methods).[18] The probe tip height moves between the bottom position, $h$, and the top-most position, $h+\Delta h$, where the values of $h$ and $\Delta h$ ($h = 10 \sim 400$ nm, $\Delta h = 0 \sim 200$ nm) are arbitrarily chosen for each measurement. Letting $I(h)$ and $I(h+\Delta h)$ be the detector signals when the tip position is at $h$ and $h+\Delta h$, we study the fundamental demodulated signal, $I_\Omega(h) = I(h) - I(h+\Delta h)$, or the second-order demodulation signal, $I_{2\Omega}(h) = I_\Omega(h) - I_\Omega(h+\Delta h)$ (see Methods). In our measurements the non-modulated detector signal $I(h)$ practically represents the far-field signal since $I_\Omega(h), I_{2\Omega}(h) << I(h)$.

In active measurements of s-SNOM, which utilizes external illumination,[19-23] it is widely known that far-field component is not filtered out in the fundamental demodulation signal, $I_\Omega(h)$, because a bulky probe shaft high above the tip scatters incident radiation. Due to this effect $I_\Omega(h)$ does not decay rapidly with increasing $h$ but extends a long tail to larger values of $h$. Accordingly, active s-SNOM measurements adopt second-order demodulation signal, $I_{2\Omega}(h)$, or still-higher order demodulation signals.[20-23] We expect that, in passive measurements as well, if the sample is heated up, the sample works as an intensified radiation source, the spontaneous emission from which is scattered by the probe shaft and causes a significant un-filtered far-field component in $I_\Omega(h)$. To avoid such complication of measurements, we do not heat up our samples but place them at ambient temperature.

**Near-field signals.** Our study includes measurements on numerous samples of patterned layers of several metals (Au, Al, and Ti) deposited on different substrates (GaAs, SiC, and SiO2). (See Methods). In this report we focus fundamental features of metals, which are found to be similar among different metals. Figures 2a shows a sample of Au layer patterned on a GaAs substrate. In the image of far-field signal $I$ without probe modulation (Fig. 2b), concentric Au rings are not resolved with the resolution, $\Delta X_{FF} \approx 15$ μm, of the confocal microscope. By modulating the probe ($h = 10$ nm, $\Delta h = 100$ nm), the detailed structure is clearly visualized in the image of near-field signal $I_\Omega$ (Fig. 2c).

Figures 2d-f displays another series of measurements on a relatively large Au disk on a SO$_2$ substrate. Near-field image ($I_\Omega$) is shown in Fig. 2d. The near-field signal ($I_\Omega$) and the far-field signal ($I$) recorded in a line scan are compared in Fig. 2e. It is noted that SiO$_2$ is brighter in the far field, but Au is brighter in the near field: The same is also true for the contrast between Au and GaAs in Fig. 2b; namely, in far fields SiO$_2$ and GaAs are brighter because the emissivities are higher than that of Au, but in the near field they are less brighter because the resonant SPhP frequencies are away from the frequency in study ($\omega = 1.30 \times 10^{14}$ rad/s).

At the boundaries of Au/GaAs (Figs. 2 c) and Au/SiO$_2$ (Fig. 2 e), the edge resolution is noted to be $\Delta X_{NF} \approx 60$ nm $\approx 0.004\ \lambda_0$, where the radius of curvature of the tip apex is found to be $R \approx 50$ nm via SEM image. Additional experiments using different probe tips ($R = 200 \sim 20$ nm) show that the resolution is roughly determined by $R$: Particularly, $\Delta X_N \approx 20$ nm is achieved with the sharpest probe tip with $R \approx 20$ nm. To examine how far the near-field evanescent waves extend out of the surface, we study $I_\Omega$ ($\Delta h = 25$ nm) as a function of $h$ on the Au disk shown in Fig. 2d. Figure 2f shows that $I_\Omega$ rapidly decreases as $h$ increases, yielding a characteristic decay length of $L_z \approx 40$ nm $\sim 0.003\ \lambda_0$. We also studied $I_{2\Omega}$ and found a similar decay length ($L_z \approx 30$ nm). We studied the decay length of $I_\Omega$ and $I_{2\Omega}$ on a number of other metal samples (Au, Al and Ti) of different shapes, and confirmed similar values: The signals practically vanish at $h \approx 150$ nm in all the samples studied. These results indicate unambiguously that the electromagnetic evanescent fields are strictly confined in the close vicinity of the surface in a range $h < 150\sim200$ nm.

In order to estimate absolute energy density of the evanescent waves, we return to Fig. 2e and note that $I_\Omega (h, \Delta h) = I(h) - I(h + \Delta h) = 440$ pA while $I(h) = 261$ nA on Au ($h = 10$ nm, $\Delta h = 100$ nm). Noting the decay profile of Fig. 2f and also considering that the near-field and the far-field components originate from the areas respectively given by $\Delta X_{NF}^2$ and $\Delta X_{FF}^2$, the respective energy fluxes emitted per unit area are, respectively, $I_\Omega/(\Delta X_{NF})^2$ and $I/(\Delta X_{FF})^2$, from which $\{I_\Omega/(\Delta X_{NF})^2\}/\{I/(\Delta X_{FF})^2\} \approx 100$ is obtained with $\Delta X_{NF} \approx 20\sim60$ nm and $\Delta X_{FF} \approx 15$ μm. Noting that the efficiency of the evanescent waves being scattered by the tip and being guided to the detector is substantially less than unity and that finite emissivity of optical components also contribute to $I$, we conclude that the energy density of the near-field evanescent waves (at $h = 10$ nm) is higher than that of the black-body by a factor much larger than 100. The decay profile and the energy density of the evanescent waves are also studied on Al and Ti, and similar results are obtained.

We note that the near-field image of the Au disk in Fig. 2d is structure-less, without exhibiting any interference patterns ascribable to SPlP-mode waves. (Considering the disk diameter and the wavelength of SPlP waves, six to seven concentric fringes should be visible in the disk if SPlPs are present.) By examining different metal patterns of varying size and shape, we are convinced that no interference pattern ascribable to SPlPs is visible. This is different from the earlier report of de Wilde et al.[12], as will be discussed later.

Two characteristic lengths, $\Delta X_{NF} \sim 20$ nm and $L_z \sim 40$ nm, of the evanescent waves strongly suggest, without invoking particular interpretation, that the evanescent electromagnetic waves ($h = 10$ nm) on metals are dominated by those of short in-plane wavelengths $\lambda_{//} < 2\pi \Delta X_{NF} \sim 0.01 \lambda_0$ and short decay lengths $l_z \sim 40$ nm $\sim 0.003 \lambda_0$; in terms of wave numbers, $k_{//} > 100 k_0$ and $ik_\perp \sim 50 k_0$.

In order to crucially test our interpretation, we study a series of metal disks with systematically varying the disk diameter $D$ from 16 μm down to 0.4 μm. The idea behind this test is straightforward: The near-field radiation intensity on a disk of $D$ will be reduced with decreasing $D$ because generation of those evanescent waves with in-plane wavelengths, $\lambda_{//}$, exceeding $D$ will be suppressed. The study is made for Au and Al disks. As shown in the inset of Fig. 3, each metal disk is separated from the substrate metal by a 100-nm-thick $Al_2O_3$ spacer layer. Figure 3 shows that the intensity of $I_\Omega$ is kept nearly unchanged with decreasing $D$ down to $D_c \sim 1$ μm $\sim 0.07 \lambda_0$ but starts decreasing rapidly below $D_c$. This feature makes certain that the evanescent waves are dominated by those of short in-plane wavelengths, $\lambda_{//} < D_c \sim 1$ μm $\sim 0.07 \lambda_0$ or large wave numbers $k_{//} > 14 k_0$, and again, rules out possible relevance of SPlP waves.

Though not shown here, we have studied temperature dependence of $I_\Omega$ on Au, Al, and Ti in a rage of 10 °C$<T<$ 60 °C and confirmed that the signal increases with increasing $T$, following a relation consistent with $I_\Omega \propto 1/\{\exp(\hbar\omega/k_B T) - 1\}$ ($\hbar$; the Dirac constant, $k_B$; the Boltzmann constant). This shows that the evanescent waves in study are thermally excited.

**Comparison with theory**
**Electromagnetic local density of states (LDOS).** All of our experimental findings described in the above are consistently accounted for by existing theories. The density of electromagnetic energy of thermally exited radiation (angular frequency $\omega$) at distance z from the surface of a semi-infinite material at temperature $T$,

$$U(z,\omega) = \rho(z,\omega) [\hbar\omega / \{\exp(\hbar\omega/k_B T) - 1\}], \quad (1)$$

has been theoretically derived, [1,13,14,24] where $\rho(z,\omega)$, called the electromagnetic local density of states (LDOS), is given by

$$\rho(z,\omega) = \rho_{PG} + \rho_{EV} = (\rho_0/2)\{\int_0^1 P_{PG}(K_{//}, z, \omega)\,dK_{//} + \int_1^\infty P_{EV}(K_{//}, z, \omega)\,dK_{//}\}, \quad (2)$$

with $\rho_0 = \omega^2/(\pi^2 c^3)$ being the LDOS in vacuum. Here $\rho_{PG}$ and $\rho_{EV}$ are the contributions from propagating and evanescent wave components, which are obtained, respectively, by integrating

$$P_{PG}(K_{//}, z, \omega) = (K_{//}/K_\perp)\{2 + K_{//}^2[\text{Re}(r_{12}^s e^{2iK_\perp \omega z/c}) + \text{Re}(r_{12}^p e^{2iK_\perp \omega z/c})]\} \quad (3)$$

and

$$P_{EV}(K_{//}, z, \omega) = (K_{//}^3/|K_\perp|)[\text{Im}(r_{12}^s) + \text{Im}(r_{12}^p)]e^{-2|K_\perp|\omega z/c}] \quad (4)$$

in the respective intervals of $K_{//}$. Here, $\mathbf{K} = (\mathbf{K}_{//}, K_\perp) = \mathbf{k}/k_0 = (\mathbf{k}_{//}, k_\perp)/k_0$ with $K_{//} = |\mathbf{K}_{//}|$ are the normalized wave vector satisfying $K_{//}^2 + K_\perp^2 = 1$, where $K_\perp$ is real for propagating waves ($0 < K_{//} < 1$) and imaginary for evanescent waves ($1 < K_{//}$). $r_{12}^s$ and $r_{12}^p$ are the Fresnel reflection coefficients for $s$- and $p$-polarizations determined by the complex dielectric constant $\varepsilon(\omega)$ of metals, which we approximate with the Drude model.[25] While each LDOS consists of the contributions from electric and magnetic components,[1,13,14] we discuss the total LDOS here.

Figure 4a displays results for Au at $\omega = 1.30 \times 10^{14}$/s. The LDOS, $\rho = \rho_{PG} + \rho_{EV}$, takes large values by more than four orders of magnitude higher than $\rho_0$ (black-body) at $z \approx 10$ nm, but rapidly decreases to a value close to $\rho_0$ at $z \approx 400$ nm, above which $\rho$ is nearly a constant at $\rho \approx \rho_0$. The contribution from evanescent waves, $\rho_{EV}$, dominates in the near-field domain ($z < 200$ nm). The nature of evanescent waves is elucidated in Fig. 4b, where two groups are distinguished. One is a group of SPlP-mode $p$-polarized surface waves, which are given by the pole of $r_{12}^p$ in Eq.(4) and form an extremely sharp peak at $K_{//} = [\varepsilon/(\varepsilon+1)]^{1/2} \approx 1.000045$ ($iK_\perp \ll 1$), as re-plotted in Fig. 4c. The other one is a group of broad-band fluctuating electromagnetic fields distributing over a wide range of extremely large values of $K_{//}$. The respective contributions to the LDOS, $\rho_{SPlP}$ and $\rho_{Fluc}$, can be separately obtained by integrating $P_{EV}(K_{//}, z)$ over $1 < K_{//} < 1.0003$ and over $1.0003 < K_{//} < \infty$. It should be noted that the SPlP peak width is so small that the integrated contribution, $\rho_{SPlP}$, is extremely small (even smaller than $\rho_{PG} \approx \rho_0$) in the entire range of z in Fig. 4a. The decay length of $\rho_{SPlP}$ is noted to be much longer than $\lambda_0$. Theory thus predicts that fluctuating electromagnetic fields of large in-plane wave numbers $K_{//}$ utterly dominate the thermal evanescent waves in the near field domain. (Large $K_{//}$ values imply also large values of $iK_\perp$ due to $K_{//}^2 + K_\perp^2 = 1$.) These theoretical predictions are consistent with our findings that (i) the near-field signal rapidly decreases with increasing the distance from the surface, and thermal evanescent waves are those of large wave numbers of $k_{//}$, $ik_\perp \gg k_0$, and (ii) no signature of SPlP waves is visible.

Let us make the analysis quantitative. We expect that the intensity of tip-scattered evanescent waves is proportional to the energy density of the evanescent waves at the tip position z, where we approximate the tip by a metal sphere of radius $R$ with its center at a height $h'$ above the probe apex ($z = h + h'$, inset of Fig. 2f). It follows that experimental demodulation signal is written as $I_\Omega(h) \propto \{|\alpha_{\text{eff}}(z)|^2 \rho(z) - |\alpha_{\text{eff}}(z+\Delta h)|^2 \rho(z+\Delta h)\}$, where $\alpha_{\text{eff}}(z) = \alpha(1+\beta) / \{1-(\alpha\beta/16\pi z^3)\}$ with $\alpha = 4\pi R^3$

($\varepsilon_W$-1)/($\varepsilon_W$+2) and $\beta = (\varepsilon-1)/(\varepsilon+1)$ is the effective polarizability of the tip.[20] Here $\varepsilon_W$ and $\varepsilon$ are the complex dielectric constants of the probe (W) and the sample metal,[25] and $R = 50$ nm is the tip radius. In Fig. 2f, theoretical values reproduce well the experimental rapid decay profile of $I_\Omega$. The reliability of analysis is assured by the parameter value $h' = 70$ nm (obtained from the best fit) being close to $R = 50$ nm.

The disk-size dependence (Fig. 3) of the near-field signal intensity is quantitatively analyzed by assuming that $\lambda_{//}$ (or the wave number) of thermal evanescent waves on the disk of diameter $D$ is cut off at $\lambda_{//} = D$ (or $K_{//} = \lambda_0/D$). Hence we evaluate the LDOS, $\rho(D)$, for the disk by integrating $P_{EV}(K_{//}, z, \omega)$ in Eq.(4) in the interval $\lambda_0/D < K_{//} < \infty$, instead of $1 < K_{//} < \infty$. As shown in Fig. 3, theoretical values of $\rho(D)/\rho(\infty)$ reproduce nicely the experimentally found size dependence for both Au and Al, supporting our interpretation. (This analysis does not include any adjustable parameter since $z = h + h' = 80$ nm with $h' = 70$ nm has been determined by the analysis of decay profile (Fig. 2f).)

**Discussion**

Our work shows that the kinetic energy of thermal motion of conduction electrons in a nano-scale region below the probe tip is efficiently converted to the energy of propagating electromagnetic waves, corresponding to an effective local emissivity more than 100 times higher than that of a blackbody. This *super emissivity* will contribute for designing a variety of novel nano-scale devices including those of near-field lithography[26] and thermophotovoltaics.[27]

While our work has demonstrated that SPlPs in the LWIR region are not efficiently excited by the stochastic thermal motion of electrons, it is interesting to ask in which condition thermal evanescent waves on metals acquire appreciable coherence. At a given frequency ω, coherent SPlP waves coexist with in-coherent fluctuating electromagnetic fields. In the LWIR region, the contribution of SPlPs to the near-field electromagnetic energy density ($\rho_{SPlP}$) is much smaller than that of the fluctuating fields ($\rho_{Fluc}$). However, the relative weight of SPlPs increases with increasing ω. In the visible region, contribution of SPlPs becomes appreciable and the long-range coherence becomes thereby explicit in the total evanescent waves.[28] Experimentally, it is possible to selectively couple the SPlP mode (in the visible or near-infrared region) to the propagating waves in far fields by using structured surfaces: This leads to highly coherent thermal sources at elevated temperatures (620K-2000K).[29-31] (Note that even in this condition, fluctuating electromagnetic evanescent fields still remain on the surface.) The situation is different in polar dielectrics, where SPhPs in the infrared dominate at room temperature.[32,33]

Finally we briefly discuss the passive s-SNOM ($\lambda_0 = 10.9 \pm 0.5$ μm) measurements of de Wilde et al., where Au samples were heated up to 170°C.[12] Very large decay lengths ($L_z$) of demodulated signals are suggested; typically, as large as $L_z \approx 3$ μm for $I_\Omega$ and $L_z \approx 200$ nm for $I_{2\Omega}$. While these results are largely different from our findings, $L_z \approx 40$ nm ($I_\Omega$) and $L_z \approx$

30 nm ($I_{2\Omega}$), they are reminiscent of active s-SNOM measurements,[19-23] in which $I_\Omega$ has a slowly decreasing long decay component due to the un-filtered far-field radiation. Interference fringes were found to appear in the images of $I_\Omega$, complicating the interpretation of results. Furthermore, the fringes were compared with theoretical electromagnetic LDOSs ρ at a height of 3 μm: The physical implication of the comparison is unclear because at z ≈3 μm the contribution from evanescent waves practically vanishes and propagating waves dominate ρ. Furthermore, our previous work gives another caution.[34] When the tip-to-surface distance exceeds ~1 μm, signals of different origin appear in $I_\Omega$: Background far-field radiation incident on the sample is reflected at the sample surface, and is also partially scattered at the probe tip, causing strong interference signals in $I_\Omega$. We are not at the position to interpret the reported fringes, but it should be mentioned that, so far as the present authors are aware, the observation of such fringe structure has not been reported again by any research groups since the first report of a decade ago.

**Methods**

**Ultra-highly sensitive CSIP detectors and home-made confocal microscope.** The energy flux of the tip-scattered near-field thermal radiation (298K) can be roughly estimated by assuming a black-body radiation emitted through a spot size $\Delta X_{NF}^2$ with $\Delta X_{NF} = 20 \sim 60$ nm being the near-field spatial resolution. By assuming a 10% spectral band width at $\lambda_0 = 14.5$ μm, we estimate the power in the 2π solid angle to be less than $10^{-13}$ Watt, which, even if reaching the detector with 100%-efficiency, is a critical level of detection when a commercial HgCdTe detector of highest sensitivity (noise equivalent power: NEP~$1.0 \times 10^{-13}$ W/Hz$^{1/2}$) is used. We overcome this difficulty by using CSIP detectors with a sensitivity (NEP~$1.0 \times 10^{-18}$ W/Hz$^{1/2}$) by orders of magnitude higher than that of best HgCdTe detectors.[16] The detection band of CSIP is relatively narrow; $\lambda_0 = 14.5 \pm 0.7$ μm for the detector used. The merit of ultra-high sensitivity of the detector is maximized by housing the detector in a cold metal with a minimized pin-hole (62 μm-ϕ) in a home-made confocal microscope (numerical aperture NA = 0.60).[17,18] Incidence of stray radiation is thereby minimized, and the high-sensitivity along with the highest spatial resolution ($\Delta X_{FF} = 0.61\lambda_0/NA \approx \lambda_0 \approx 15$ μm; theoretical diffraction limit) are achieved.

**Custom-made AFM system with independent modulation of probe tip height.** Extracting near-field signal component out of far-field background component is crucially important because the far-field background radiation is much stronger. The relevant sample area for the far-field background radiation is the focus spot of the far-field geometrical optics, $\Delta X_{FF}^2$, which is by orders of magnitude larger than the area, $\Delta X_{NF}^2$, relevant to the focus spot of near-field optics. The ratio $\Delta X_{FF}^2/\Delta X_{NF}^2$ is around $10^5$ when $\Delta X_{FF} \sim \lambda_0 \approx 15$ μm and $\Delta X_{NF} = 20 \sim 60$ nm are used. In actual measurements, the far-field background component is typically about $10^3 \sim 10^4$ of the near-field signal component.

The probe tip position is controlled in the shear-force mode of AFM, where the tip vibrates parallel to the surface at a frequency of $f \approx 32$ kHz in a small amplitude ~2 nm.[18] At the same time, the tip is moved up and down at a frequency $\Omega = 10$ Hz between the heights $h$ and $h + \Delta h$ by using an additional piezo actuator. In real practice, the bottom height $h$ in each cycle is monitored/controlled instantaneously (with a time constant ~3 ms) in the shear-force mode AFM control. The values of $h$ and $\Delta h$ can be chosen as independent experimental parameters in ranges of $h = 10 \sim 400$ nm and $\Delta h = 0 \sim 200$ nm. Typically, the atomic force is appreciable only in a range of $h < 40$ nm. So we additionally define a reference set point at $h^* = 10$ nm for detecting the probe height to determine $h$ and $h + \Delta h$, when we choose 40 nm $< h$.

**Demodulation signals.** The demodulating signal, $I_\Omega(h) = I(h) - I(h + \Delta h)$ or $I_{2\Omega}(h) = I_\Omega(h) - I_\Omega(h + \Delta h)$, is obtained either by taking differential signal or by demodulating the signal $I(h)$ with a Lock-in amplifier at fundamental frequency $\Omega$ or at second harmonic frequency $2\Omega$. A strong evidence showing that the near-field evanescent wave is strictly extracted in our measurements is that $I_\Omega(h)$ decays rapidly with increasing $h$ ($L_z \approx 40$ nm) in substantially the same manner as that of $I_{2\Omega}(h)$ ($L_z \approx 30$ nm). Furthermore, by thoroughly studying $I_\Omega(h)$ and $I_{2\Omega}(h)$ as a function of both $h$ and $\Delta h$, we can confirm the consistency of our measurements. We are thereby convinced that extraction of the near-field component is perfect both in $I_\Omega(h)$ and $I_{2\Omega}(h)$.

**Fabrication of samples.** The metal layers are patterned to a thickness of 50-80 nm on the substrates via standard optical and electron-beam lithography techniques. The metal layers are thick enough to avoid substrate-specific effects.[35]

**Acknowledgements**

This work was supported by CREST project and **JST-SENTAN program** of Japan Science and Technology Agency (JST). S.K thanks Shanghai Institute of Technical Physics (SITP), Chinese Academy of Sciences (CAS) for the support under the Special Program of Foreign Experts.

**Author contributions**

S.K. planned and led the project, designed the detectors, supported experiments and data analysis, and wrote the manuscript. Y.K. performed experiments and data analysis, and wrote the manuscript. K.K. supported experiments and data analysis. Z.A. and T.U. designed and fabricated the detectors. All authors contributed to writing the manuscript.

**Additional information**

Supplementary information is available in the online version of the paper. Correspondence and requests for materials should be addressed to S.K.

**Competing financial interests**

The authors declare no competing financial interests.

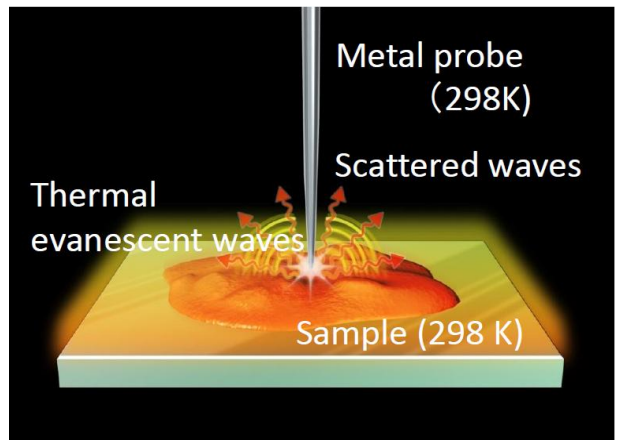

# Figure 1

**Passive near field microscopy.**

Electromagnetic evanescent waves thermally excited in the close vicinity of material surface are scattered by a sharp metal probe tip (W), and the scattered waves are collected by a long-wavelength infrared (LWIR) confocal microscope and detected with an ultra-highly sensitive CSIP detector. The detected wavelength is $\lambda_0 = 14.5 \pm 0.7$ μm ($\omega = 2\pi c/\lambda_0 \approx 1.30 \times 10^{14}$ rad/s). See text for detailed description.

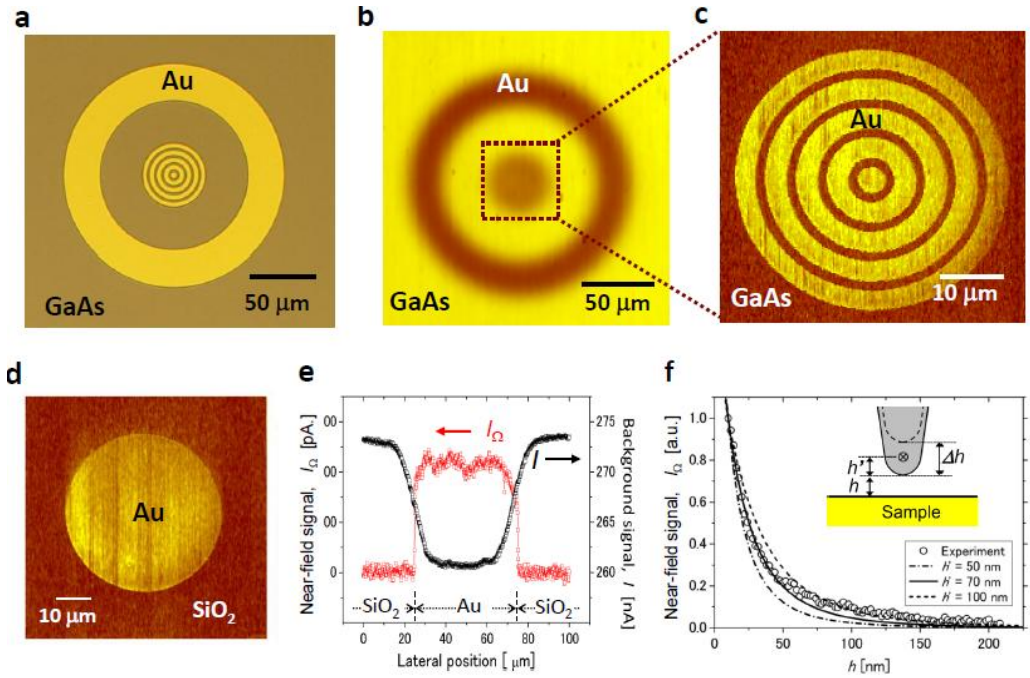

## Figure 2

**Near-field signals at room temperature.**

**a**, Optical microphotograph of a sample with a pattern of 80 nm thick Au layer deposited on a GaAs substrate. The Au pattern consists of a large outer ring with smaller concentric 3 μm-wide (5 μm-pitch) rings in the central region. **b**, Far-field image ($I$) of the sample shown in **a**, obtained without probe modulation ($\lambda_0 = 14.5 \pm 0.7$ μm). **c**, Near-field image ($I_\Omega$) obtained with $h = 10$ nm and $\Delta h = 100$ nm. **d**, Near-field image ($I_\Omega$) with $h = 10$ nm and $\Delta h = 100$ nm of a 50 μm-diameter Au disk on a SiO$_2$ substrate. **e**, Comparison of the far-field background signal, $I$, and the near-field signal, $I_\Omega$ ($h = 10$ nm, $\Delta h = 100$ nm), simultaneously recorded in the linear scan on the sample shown in **d**. The edge resolution at the Au/SiO$_2$ boundary is $\Delta X_{FF} \approx 15$ μm for the far-field signal and $\Delta X_{NF} \approx 60$ nm for the near-field signal. **f**, Near-field signal $I_\Omega (h) = I(h) - I(h + \Delta h)$ with $h = 10$ nm and $\Delta h = 25$ nm (open circles) as a function of $h$ on the sample shown in **d**. Three curves indicate theoretical values obtained by assuming $h' = 50$, 70 and 100 nm with $h = 10$ nm and $\Delta h = 25$ nm (see text).

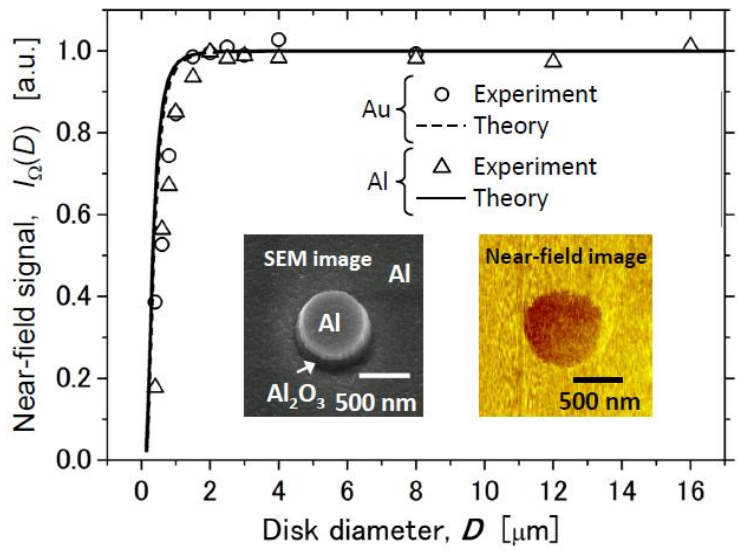

## Figur 3

**Sample-size dependence of near-field signals.**

Near-field signal $I_\Omega$ ($h$ = 10 nm, $\Delta h$ = 200 nm) versus the diameter $D$ of Al and Au disks, where the signal intensity is normalized by the values of large disks ($D > 8$ μm). The solid line and the broken line indicate, respectively, theoretical values of normalized LDOS, $\rho(D)/\rho(\infty)$, for disks of Al and Au with $z$ = 80 nm ($h$ = 10 nm, $h'$ = 70 nm). The left inset is a scanning electron microscope (SEM) image of an Al disk of $D$ = 600 nm. The near-field image ($I_\Omega$) on the right shows that the Al disk is darker than the background Al plate.

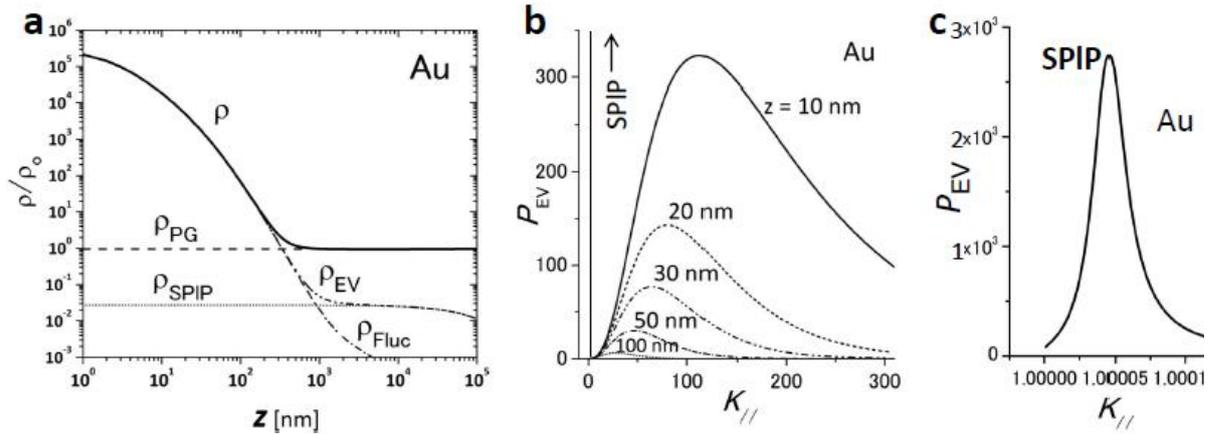

Figure 4

**Electromagnetic local density of states (LDOS).**

**a**, LDOS versus distance $z$ from the surface of Au at $\omega=1.30 \times 10^{14}$/s ($\lambda_0 = 14.5$ μm). The values are normalized by the value in vacuum $\rho_0 = \omega^2/(\pi^2 c^3)$. The total LDOS, $\rho = \rho_{EV} + \rho_{PG}$, includes both of the contributions from evanescent waves, $\rho_{EV}$ and propagating waves, $\rho_{PG}$, but $\rho \approx \rho_{EV}$ in the near field ($z < 200$ nm) because $\rho_{EV}$ is much larger than $\rho_{PG}$. The contribution of evanescent waves, $\rho_{EV} = \rho_{SPIP} + \rho_{Fluc}$, consists of the components of SPlP-mode $p$-polarized surface waves, $\rho_{SPIP}$, and short-wavelength fluctuating electromagnetic fields, $\rho_{Fluc}$. The contribution from SPlP waves, $\rho_{SPIP}$, however, is very small (even smaller than $\rho_{PG}$), so that the total LDOS is utterly dominated by the fluctuating fields in the near field; viz, $\rho \approx \rho_{Fluc}$.  **b**, $P_{EV}(K_{//}, z)$ versus $K_{//} = k_{//}/k_0$. Two distinct components are distinguished. One is the sharp peak located at $K_{//} \approx 1.000045$, which is the manifestation of SPlP-mode $p$-polarized waves. Whereas the SPlP peak is high (a peak value reaching ca. 2600), its contribution to the LDOS, $\rho_{SPIP}$, is negligibly small because its integrated intensity is small. Another component, providing the dominant contribution, $\rho_{Fluc}$, forms a broad band of short-wavelength fluctuating fields covering a wide range of large $K_{//}$-values: The $K_{//}$-range of distribution along with the amplitude rapidly increases as z decreases. **c,** Re-plot of the SPlP-peak. Unlike the broad band of fluctuating fields, the SPlP peak (both the width and the height) is confirmed to be substantially unchanged as z varies in a range of 1nm< z <1000 nm.